\documentclass[aps,pra,twocolumn,showpacs,notitlepage]{revtex4-1}

\usepackage{graphicx}

\begin{document}

\title{Reasonable conditions for joint probabilities of non-commuting observables}

\author{Holger F. Hofmann}
\email{hofmann@hiroshima-u.ac.jp}
\affiliation{
Graduate School of Advanced Sciences of Matter, Hiroshima University,
Kagamiyama 1-3-1, Higashi Hiroshima 739-8530, Japan}
\affiliation{JST, CREST, Sanbancho 5, Chiyoda-ku, Tokyo 102-0075, Japan
}

\begin{abstract}
In the operator formalism of quantum mechanics, the density operator describes the complete statistics of a quantum state in terms of $d^2$ independent elements, where $d$ is the number of possible outcomes for a precise measurement of an observable. In principle, it is therefore possible to express the density operator by a joint probability of two observables that cannot actually be measured jointly because they do not have any common eigenstates. However, such joint probabilities do not refer to an actual measurement outcome, so their definition cannot be based on a set of possible events. Here, I consider the criteria that could specify a unique mathematical form of joint probabilities in the quantum formalism. It is shown that a reasonable set of conditions results in the definition of joint probabilities by ordered products of the corresponding projection operators. It is pointed out that this joint probability corresponds to the quasi probabilities that have recently been observed experimentally in weak measurements.
\end{abstract}


\maketitle
\section{Introduction}
\label{sec:intro}
Quantum mechanics is a statistical theory that defines the probability of measurement outcomes without referring to a fundamental set of possible realities. The original formulation of the theory was based on analogies between the algebra of operators and the algebra of numbers. However, this analogy is somewhat misleading, since individual measurement outcomes are not described by the operators but by their eigenvalues. As a consequence, there is no quantum mechanical equivalent to a phase space point $(x,p)$ because position and momentum do not have common eigenstates. Nevertheless classical mechanics should emerge as a valid approximation of quantum statistics, so it would seem natural to ask how the notion of phase space points can emerge from a theory that does not assign any joint reality to $x$ and $p$. 

Early attempts to describe the relation between classical phase space statistics and quantum statistics focussed on formal relations that apply specifically to continuous variables and the Fourier transform relation between the eigenstates of position and momentum. Specifically, Wigner showed that the classical phase space distribution could be approximated by a Fourier transform along the anti-diagonal of the spatial density matrix, resulting in a quasi probability expression for the density operator that is now widely known as the Wigner function \cite{Wig32}. Almost immediately after this historic result, Kirkwood pointed out that a similar analogy with classical phase space distributions could be obtained by a more simple Fourier transform applied to only one side of the density matrix \cite{Kir33}. This quasi probability is necessarily complex, but it converges on the same classical limit and also produces the correct marginal distributions for both position and momentum. The early history of quasi probabilities thus illustrates the problem of finding a unique definition of joint probabilities in the absence of actual joint measurements.

Recent developments in quantum information have seen a more general discussion of quantum mechanics as a statistical theory \cite{Har01,Fuc13}. In the spirit of these discussions, it may be worthwhile to reconsider the concept of joint probability based on the general operator algebra of quantum statistics. Specifically, it may be possible to derive a definition of joint probability from a set of reasonable conditions or axioms that characterize the relation between the joint probabilities and the actual measurement results. In the following, I propose a set of axioms that results in a definition of joint probability which is consistent with the quasi probability introduced by Kirkwood and therefore provides an objective reason for excluding the Wigner function. The essential criterion that eliminates alternative definitions of joint probabilities concerns the relation between physical properties with joint eigenstates: to ensure that the probabilities of outcomes associated with the same joint eigenstate of the two properties are the same in both measurements, the joint probabilities must be defined by a product of projectors that eliminates all states orthogonal to either of the two eigenstates. For all other definitions of joint probability, there will be non-zero joint probabilities for properties that directly contradict the known properties of the input state. It is therefore possible to argue that the product of projection operators is the only valid representation of a logical AND in the quantum formalism, resulting in the definition of a complex valued joint probability that is unique except for the ordering dependent sign of its imaginary part.

\section{The operator algebra of joint probabilities}
\label{sec:jprob}

The motivation for a definition of joint probabilities of non-commuting observables can be explained in terms of the calculation of probabilities in the Hilbert space formalism. In Hilbert space, a state is represented by a d-dimensional complex vector, where the absolute values of the vector components represent the probabilities of measurement outcomes. However, the outcomes of other measurements will depend on the differences between the complex phases of the $d$ components. In the density matrix, these complex phases appear in the off-diagonal elements. In general, the probability of a measurement outcome $m$ is therefore given by a sum over all matrix elements of the density operator $\hat{\rho}$ and the measurement operator $\hat{\Pi}(m)$, as given by the product trace
\begin{eqnarray}
\label{eq:Pm1}
P(m) &=& \mbox{Tr}\left( \hat{\Pi}(m) \hat{\rho} \right)
\nonumber \\ &=&
\sum_{a,a^\prime} \langle a \mid \hat{\Pi}(m) \mid a^\prime \rangle \langle a^\prime \mid \hat{\rho} \mid a \rangle. 
\end{eqnarray}
If $a$ and $a^\prime$ refered to different properties, one could identify the matrix elements of $\hat{\rho}$ with joint probabilities and the matrix elements of $\hat{\Pi}(m)$ with conditional probabilities, and this analogy is probably behing the somewhat irritating claim that superposition assigns simultaneous reality to different and distinct values of the same property (the particle is ``simultaneously'' here and there, or the cat is ``both'' dead and alive). However, the off-diagonal elements do not appear in the measurement statistics of $a$ at all - they are only relevant for measurements of a different property $b$. It would therefore seem natural to express the density operator in terms of a joint probability of $a$ and $b$, so that general measurement probabilities could be expressed in closer analogy to classical statistics as 
\begin{equation}
\label{eq:Pm2}
P(m) = \sum_{a,b} P(m|a,b) \rho(a,b).
\end{equation}
Note that the number of matrix elements and the number of joint probabilities is both given by the square of the Hilbert space dimension $d^2$. Thus, the algebra of Hilbert space matrices is very similar to the algebra of joint and conditional probabilities. All it takes to make the connection is a transformation of the matrix representation into a joint probability representation. In general, this transformation can be represented by an operator $\hat{\Pi}(a,b)$ that assigns a joint probability $\rho(a,b)$ to the density opertor $\hat{\rho}$ through the product trace,
\begin{equation}
\rho(a,b) = \mbox{Tr}\left( \hat{\Pi}(a,b) \hat{\rho} \right).
\end{equation} 
The construction of the operator $\hat{\Pi}(a,b)$ defines the joint probabilities $\rho(a,b)$. However, a meaningful definition of joint probabilities must satisfy a number of criteria that motivate the specific choice of $\hat{\Pi}(a,b)$ in terms of reasonable assumptions about the relation between the projective measurements of $a$ and $b$. In the following, I will formulate such a set of reasonable assumptions and show that they narrow down the mathematical possibilities for a definition of $\hat{\Pi}(a,b)$ to products of the projection operators. 

\section{Reasonable requirements}
\label{sec:cond}

The first obvious requirement of joint probabilities is that they should correctly describe the individual probabilities of $a$ and of $b$ observed in separate measurements of the two observables. Since the measurement operators of these measurements are given by the projectors of $a$ and of $b$, this condition can be applied directly to the operator algebra of $\hat{\Pi}(a,b)$.

\newtheorem{cond}{Condition}
\begin{cond}
The marginals of the joint probabilities correspond to the probabilities of separate measurements of $a$ and $b$,
\begin{eqnarray}
\sum_{b} \hat{\Pi}(a,b) &=& \mid a \rangle \langle a \mid,
\nonumber \\
\sum_{a} \hat{\Pi}(a,b) &=& \mid b \rangle \langle b \mid.
\end{eqnarray}
\end{cond}

Next, it is useful to consider a situation where we have some confidence about the correct joint probability - specifically, the case where the input state $\hat{\rho}$ is an eigenstate of one of the observables with an eigenvalue of $a_\psi$ or $b_\psi$. In that case, it is reasonable to assume that the joint probabilities are zero for all other values of $a$ or $b$, so that the joint probability is given by the marginal probabilities $|\langle a \mid b \rangle|^2$.

\begin{cond}
Joint probabilities for input states with a precisely known value of $a$ or $b$ are zero for any other value of that obserable,
\begin{eqnarray}
\langle a_\psi \mid \hat{\Pi}(a,b) \mid a_\psi \rangle &=& \delta_{a,a_\psi}|\langle a \mid b \rangle|^2,
\nonumber \\
\langle b_\psi \mid \hat{\Pi}(a,b) \mid b_\psi \rangle &=& \delta_{b,b_\psi}|\langle a \mid b \rangle|^2.
\end{eqnarray}
\end{cond}

It may seem that this requirement is rather trivial, but it does eliminate all contributions to $\hat{\Pi}(a,b)$ that never show up in the marginal probabilities of $a$ or of $b$ because the sums over either $a$ or $b$ are all zero. It is rather easy to construct such artifacts, e.g. by adding and subtracting an arbitrary operator to each $\hat{\Pi}(a,b)$, so that there are equal numbers of additions and subtractions in each line or column defined by constant $a$ or $b$. Effectively, these constructions will introduce correlations into the joint probabilities even when one of the properties does not have any fluctuations that could be correlated to the other property. Thus, condition 2 could be summarized as ``no correlation without fluctuation''.

Importantly, the second condition refers only to the specific sets of outcomes $\{a\}$ and $\{b\}$ that define the complete probability distribution. It is possible to formulate a more general condition that actually includes the second condition as a specific case by considering possible superpositions of a finite subset of $a$ ($b$). In this case, the input state $\mid m \rangle$ can be distinguished from the eigenstates of $a$ ($b$) by a projective measurement on a different property that has both $\mid a \rangle$ ($\mid b \rangle$) and $\mid m \rangle$ as eigenstates. We can therefore conclude that knowledge of $m$ excludes the possibility of $a$ ($b$) in the same way that the knowledge of $a_\psi$ excluded the possibilities of other values of $a$. 

\begin{cond}
If the input state is characterized by the eigenvalue $m$ of a property that has a joint measurement outcome $m(a)$ ($m(b)$) with $a$ ($b$) which distinguishes $a$ ($b$) from the input $m$, then the joint probabilities for this measurement outcome $a$ ($b$) must all be zero.  
\begin{eqnarray}
\label{eq:c3}
\langle m \mid \hat{\Pi}(a,b) \mid m \rangle = 0 &\; \mbox{\rm if} \;& |\langle a \mid m \rangle|^2=0,
\nonumber \\
\langle m \mid \hat{\Pi}(a,b) \mid m \rangle = 0 &\;\mbox{\rm if}\;& |\langle b \mid m \rangle|^2=0.
\end{eqnarray}
\end{cond}

This condition eliminates the possibility that positive and negative joint probabilities for a specific outcome average to zero in the sums that determine the marginal probabilities. Whenever a marginal probability of zero is observed, the joint probabilities for this marginal must all be zero. Note that the reason for this condition relies on the obsevation that orthogonality of states implies that the states represent different outcomes of the same measurement. If the marginal probability of $a$ is zero, there is a direct experimentally observable contradiction between $a$ and the initial condition $m$, so that $m(a) \neq m$.

Significantly, the third condition is violated by the Wigner function, since the Wigner function associates coherences between $x$ and $x^\prime$ with the average position of $(x-x)^\prime/2$, which can have a marginal probability of zero. For example, the Wigner function of a particle passing through a double slit has non-zero values at the position between the two slits, where there is not even an opening for the particle to pass through the screen. Thus, despite its usefulness in the evaluation of measurement statistics, the value of the Wigner function for a specific combination of $x$ and $p$ does not originate from the possibility of finding the position $x$ or the momentum $p$ in independent measurements. 

In general, the third condition is necessary in order to satisfy the expectation that the joint probability of $a$ and $b$ establishes a relation between measurement results that can actually be observed in separate measurements of $a$ and of $b$. Although it is mathematically possible to define joint functions of the quantities $a$ and $b$ that do not satisfy this condition, such functions do not express any relation between the individual outcomes $a$ and $b$ and should therefore not be considered joint probabilities. Since the values of the Wigner function at $x$ can be traced to a quantitative average of pairs of outcomes other than $x$, it does not actually qualify as a joint probability of the single outcome $x$ and the single outcome $p$.

We can now apply the requirements and find the specific definition of $\hat{\Pi}(a,b)$ that satisfies all of them. In particular, the third requirement greatly reduces the number of possibilities. Since Eq.(\ref{eq:c3}) applies to all possible states $\mid m \rangle$, the operator $\hat{\Pi}(a,b)$ must assign a value of zero to any state that is orthogonal to either $\mid a \rangle$ or $\mid b \rangle$. Since such an assignment of zero is only possible by multiplication with the corresponding projection operator, the third condition can only be satisfied if the operator $\hat{\Pi}(a,b)$ is given by a product of the two projection operators. According to condition 1, there can be no additional factors, too. Only the choice of the operator ordering is arbitrary. In general, it is possible to chose any linear combination of the two orderings, but the choice of a specific ordering greatly simplifies the mathematical properties of the expression. If the projection on $a$ is applied first, the operator defining the joint probabilities reads
\begin{equation}
\label{eq:piform}
\hat{\Pi}(a,b) = \mid b \rangle \langle b \mid a \rangle \langle a \mid.
\end{equation}
Since the eigenvalues of the projection operators represent the truth values of the statements associated with their state vectore, the product of two projectors corresponds to the classical definition of a logical AND as the product of two truth values. The definition of joint probabilities using the product of the projection operators is therefore consistent with the original idea that numbers should be replaced by operators. However, the replacement of truth values with projection operators has non-trivial consequences, since the non-commutativity of the two projection operators results in a non-hermitian operator that cannot be interpreted as a projector onto a joint reality of $a$ and $b$. Instead, the quantum mechanical relation between the separate realities of $a$ and $b$ is expressed by a complex valued joint probability obtained from the expectation values of the non-hermitian operator $\hat{\Pi}(a,b)$. In the following, I will point out that complex probabilities of this kind have a long history in quantum physics, perhaps culminating in the realization that they can be obtained experimentally in weak measurements. It is then possible to explain the physics expressed by the operator ordering and to consider wider implications for the foundations of quantum physics. 

\section{Joint probablities in quantum physics}
\label{sec:phys}

The discussion above is based entirely on the structure of the Hilbert space formalism and on conditions derived from projective measurements of operator eigenvalues. In particular, it was not based on methods of quantum state reconstruction by tomographically complete sets of measurements, which have often been used as a motivation for the introduction of joint probabilities. It is interesting to note that an expression for joint probabilities can be derived without any reference to joint measurements, only by considering the structure of the operator formalism and its application to separate projective measurements of $a$ and $b$. 

Since the result given in Eq.(\ref{eq:piform}) is a simple multiplication of projection operators, it appears in the equations of the operator algebra whenever two operators with eigenstates $\{\mid a \rangle\}$ and $\{\mid b \rangle\}$ are multiplied. It is therefore not surprising that the joint probability defined by Eq.(\ref{eq:piform}) has already been studied in other contexts. As mentioned above, its application to position and momentum results in the distribution introduced by Kirkwood in 1933 \cite{Kir33}. The general form for arbitrary pairs of observables was introduced by Dirac in 1945 \cite{Dir45}. These early works have recently attracted renewed attention, since it was discovered that the complex joint probabilities of Kirkwood and Dirac actually describe the results of weak measurements of a projection operator $\mid a \rangle\langle a \mid$ followed by a final measurement of $\mid b \rangle$ \cite{Joh07,Hof12,Lun12,Wu13,Bam14}. Complex joint probabilities therefore have a well-defined operational meaning that directly relates them to sequential measurements of the two non-commuting obervables. It is also significant that the complex joint probabilities completely characterize quantum states and processes. They can therefore be used as a starting point for a fundamental reformulation of quantum physics based on empirical principles \cite{Hof14}.

In the present context, it is interesting to note that the relation with weak measurement also explains the dependence of $\hat{\Pi}(a,b)$ on operator ordering: the imaginary part of the weak value actually represents the response of the system to the dynamics generated by the observable \cite{Hof11,Dre12}. Upon time reversal, the direction of the force is inverted and the response changes its sign. It is therefore possible to identify the particular ordering with a temporal sequence and the sign of the imaginary part as the direction of the dynamics generated by the observables.  

In the formal sense, a specific operator ordering is desirable because it is mathematically convenient. As Kirkwood already noticed in 1933, the joint probability defined by $\hat{\Pi}(a,b)$ simply corresponds to the application of different basis sets to the right and the left side of the density matrix, 
\begin{equation}
\rho(a,b) = \langle b \mid a \rangle \langle a \mid \hat{\rho} \mid b \rangle.
\end{equation}
The relation between $\rho(a,b)$ and a measurement probablity $P(m)$ is then naturally expressed in the form given by Eq.(\ref{eq:Pm2}), where 
\begin{equation}
P(m|a,b) = \frac{\langle b \mid \hat{\Pi}(m) \mid a \rangle}{\langle b \mid a \rangle}.
\end{equation}
This complex conditional probability happens to be the weak value of the measurement operator $\hat{\Pi}(m)$ for an input state $\mid a \rangle$ and a post-selected state $\mid b \rangle$. It is therefore possible to obtain its value experimentally by a weak measurement of the fundamental relation between the physical properties $a$, $b$, and $m$. Since this relation can be applied to any quantum state $\hat{\rho}$, it actually describes the deterministic relation between the properties $(a,b)$ and $m$ \cite{Hof12}. Thus, complex valued conditional probabilities take the place of analytical functions that relate the values of physical properties to each other. Complex conditional probabilities actually represent the most fundamental formulation of the laws of physics, universally valid in both the quantum and the classical regime. It is therefore no accident that the quantum formalism results in a very specific definition of joint probabilities: what seemed to be ambiguities in the physics described by the operator algebra are actually well defined differences between the unjustified expectation of joint realities and the correct relations between different potential realities that is observed in sufficiently precise experiments \cite{Hof14}.

\section{Conclusions}
\label{sec:concl}

The analysis above has shown that a relatively small set of reasonable assumptions can narrow down the possible definitions of joint probabilities for two non-commuting observables to the complex joint probabilities obtained from products of the two projection operators. Any other definition of joint probabilities would introduce non-zero probabilities for events that are never observed under the conditions described by the quantum state in question. 

It seems to be significant that no other quasi probabilities can satisfy these simple requirements. The conclusion appears to be that the standard formalism of quantum mechanics is much more specific regarding the precise relations between non-commuting properties than the conventional textbook discussions of uncertainty and superpositions suggest. Ultimately, the complex joint probabilities obtained by simply multiplying the projection operators and taking the product trace with the density matrix provide an explanation of quantum effects that avoids many of the ambiguities associated with the Hilbert space formulation and may therefore help to clarify the origin of quantum paradoxes and other failures of classical explanations in quantum physics.

\section*{Acknowledgment}
This work was supported by JSPS KAKENHI Grant Number 24540427.

\end{document}